\documentclass[3p,preprint,12pt]{elsarticle}
\usepackage{graphics}
\usepackage{amssymb}

\journal{Nuclear Physics A}

\begin{document}

\begin{frontmatter}

\title{Distributions of the $S$-matrix poles in Woods-Saxon and cut-off Woods-Saxon  potentials}

\author[label1]{P. Salamon}
\author[label2]{\'A. Baran}
\author[label1,label2]{T. Vertse\corref{author}}

\address[label1]{MTA Institute for Nuclear Research, Debrecen, PO Box 51, H--4001, Hungary}
\address[label2]{Faculty of Informatics, University of Debrecen, PO Box 400, H--4002 Debrecen, Hungary}

\cortext[author] {Corresponding author\\\textit{Tel.:} +3652509231;\ \textit{fax:} +3652416181\\\textit{E-mail address:} vertse.tamas@atomki.mta.hu}

\begin{abstract}
The positions of the $l=0$  $S$-matrix poles  are calculated 
 in generalized Woods-Saxon (GWS) potential and in cut-off generalized Woods-Saxon (CGWS) potential. The solutions of the radial equations are calculated numerically for the CGWS potential and analytically for GWS using the formalism
of Gy. Bencze \cite{[Be66]}. We calculate CGWS and GWS cases at small non-zero values of the diffuseness in order to approach the square well potential and to be able to separate effects of the radius parameter and the cut-off radius parameter. In the case of the GWS
potential the wave functions are reflected at the nuclear radius therefore the
distances of the resonant poles depend on the radius parameter of the potential. In CGWS potential the wave function can be reflected at larger distance where the potential is cut to zero and the derivative of the potential does not exist.  The positions of most of the resonant poles do depend strongly on the cut-off radius of the potential, which is an unphysical parameter. Only the positions of the few narrow resonances in potentials with barrier are not sensitive to the cut-off distance.
For the broad resonances 
the effect of the cut-off can not be corrected by using a suggested analytical form of the first order perturbation
correction.
\end{abstract}


\end{frontmatter}
\vskip 0.5 cm

\section{Introduction}
Exactly solvable quantum mechanical models proved to be invaluable tools in 
the understanding of many fundamental quantum mechanical concepts. In particular, 
they give insight into complex phenomena, like the symmetries of quantum 
mechanical systems, and they also allow the investigation of transitions through critical parameter domains \cite{[Le15]}. 
Our goal in this paper is more simple, and analytical solution serves as a firm basis of  numerical integration of the radial Schroedinger equation. 

In this work we consider a very simple case in which a  neutral particle
is scattered on a spherically symmetric target nucleus represented
by a  real Woods-Saxon (WS) type \cite{[Wo54]} nuclear potential $v(r)$.
The WS potential is the most common phenomenological nuclear potential used
in the description of nuclear reactions in the Gamow shell model (GSM) description  \cite{[Mi09]} of drip line nuclei. A recent analysis of this type is the description of the  $^7$Be(p,$\gamma$)$^8$B and the $^7$Li(n,$\gamma$)$^8$Li reactions in Ref. \cite{[Fo15]}.
The key elements of GSM are the Berggren-ensembles of single particle
states. The   Berggren-ensemble might include resonant and sometime anti-bound
states beside the bound states and the scattering states along a complex path.
The shape of the path determines the set of the $S$-matrix pole states to be included in the ensemble. In order to have smooth contribution from the scattering states the shape of the path should go reasonably far from the poles. Therefore to know, where the poles are located has crucial importance.

The WS potential was introduced long time ago as a smoothed replacement of the square well (SQ)
potential, because the sudden jump of the constant depth of the potential to zero was considered unrealistic in the SQ potential. An introduction of a diffuseness parameter simulated the gradual decrease of the potential value
in the surface region of the nucleus. The only disadvantage of the diffuse potential was that the closed analytical form of the solution of the radial Schroedinger
equation had to be sacrificed. More precisely, for zero angular momentum the
solution could be calculated  analytically but for higher partial waves the equation could be solved without approximation \cite{[Ku15]} only by using numerical integration.
 In this work we use the opportunity of having both analytic and numeric solutions for the $l=0$ case and compare the 
distribution of the 
 poles of the scattering matrix $S$ for the two types
of the potentials.

In section 2 of the paper we define the poles of the S-matrix on the complex $k$-plane for zero angular 
momentum.
In section 3 we give analytical formulae of the $S$-matrix for WS and GWS potentials. In section 4 and 5 we
present the cut-off versions of the potentials (CWS and CGWS potentials) and the numerical methods for
calculating the positions of the resonant poles and the normalization of the resonant wave functions. A simple perturbation correction of the cut-off of these
potentials is also given in the section 5. Section 6 contains the results of the numerical
examples for heavy and light nuclear systems. The summary of the results is given in section 7.

\section{Poles of the $S$-matrix on the complex $k$ plane}

The nuclear potential $v(r)$ determines the so called {\it squared local wave number}

\begin{equation}
\label{locwn}
k^2(r)=[k^2 -v(r)]~,
\end{equation}
and the radial Schroedinger equation to be solved is
\begin{equation}
\label{radial}
u^{\prime\prime}(r,k)+k^2(r) u(r,k)=0~.
\end{equation}
Here prime denotes the derivative with respect to (wrt) the radial distance $r$.

 The first boundary condition (BC) for the solution $u(r,k)$ is its regularity at $r=0$: 
\begin{equation}
\label{regular}
u(0,k)=0~.
\end{equation}

The other BC is specified
 in the asymptotic region, where the nuclear potential becomes zero, and
 the solution $u(r,k)$
is proportional to a combination of the incoming $e^{-ikr}$ and outgoing $e^{ikr}$ 
free spherical  waves: 
 \begin{equation} 
\label{scattbc}
u(r,k)=C [e^{-ikr}-S_{l=0}(k)e^{ikr}]~.
\end{equation}
The ratio of the two type of solutions is fixed by
 the element of the scattering matrix $S_0(k)$. For scattering solutions 
 the BC in Eqs. (\ref{regular}) and (\ref{scattbc}) can be satisfied for any value of $k$. 
 
For the poles of $S(k)$ 
the BC should be of purely outgoing type. For a decaying resonance the solution  in the asymptotic region is
proportional to
$e^{ikr}$.
 The poles of $S_0(k)$ for a real potential are either real energy poles (bound and anti-bound poles) or complex energy poles. The real energy poles lie on the imaginary $k$-axis. The complex energy poles are the resonances.
 The decaying resonances lie in the fourth quadrant of the $k$ plane. The mirror images 
of the decaying resonances 
 are the capturing resonances in the third quadrant.
 The complex wave number of a decaying (Gamow or Gamow-Siegert) resonance is denoted by $k=k^R+ i k^I$ with $k^I<0$, and $k^R>0$, and by $k^R<0$ for a capturing resonance. Since the energy is proportional to $k^2$, therefore the unbound poles (i.e. the anti-bound poles and the resonances) lie on the second  energy-sheet. 
 
 \section{WS and GWS potentials}
 
 We want to study the dependence of the resonant pole positions on the tail of the nuclear potential. We compare the pole structure of the potentials  without any cut
and  that of the cut-off potentials. A SQ potential is zero beyond its radius
but the WS potential becomes zero only at infinity. It is convenient to write
the WS potential as a product of its strength $V_1$
\begin{equation}
\label{WS}
V^{WS}(r)=V_1f^{WS}(r)~,
\end{equation}
and
its radial shape:
\begin{equation}
\label{WSRS}
f^{WS}(r)=-
\frac{1}{1+e^{\frac{r-R}{a}}}~.
\end{equation}
This shape has two parameters, the radius $R$ and the diffuseness $a$.  The exponential tail in Eq. (\ref{WS})
tends to zero when
the radial distance $r$ goes to infinity. To calculate scattering cross-sections
we need the matrix element of the scattering matrix $S$. The value of $S_0$ can be calculated from matching
the solution of the radial Schroedinger equation to that of the  asymptotical solution.
 For a WS potential the matching to the asymptotic solutions  can be
done only at infinite distance. 
Sometimes we can complement the WS potential with a surface type term with radial shape

\begin{equation}
\label{surfaceform}
f^{SWS} (r)=-\frac{e^{\frac{r-R}{a}}}{(1+e^{\frac{r-R}{a}} )^2}~,
\end{equation}
and strength $V_2$ and consider the 
 so called generalized WS potential (GWS)
\begin{equation}
\label{general}
V^{GWS}(r)=V_1f^{WS} (r)+V_2f^{SWS} (r)~.
\end{equation}

 For
$l=0$ the solution is given
analytically by Gy. Bencze \cite{[Be66]} and very recently by O. Bayrak and E. Aciksoz \cite{[Ba15]}. Thanks to the analytical solution the matching to 
the free solution
 can be performed at infinity.
For GWS potential the $l=0$ scattering matrix element $S_0(k)$ at the complex wave number $k$ is given in Ref. \cite{[Be66]} as follows:

\begin{equation}
\label{sform}
S_0(k)=\exp(-2ikR){\Gamma(2ika)\over \Gamma(-2ika)}\cdot
{A{\Gamma(1-2\lambda)\over \Gamma(1-\lambda-\mu+ika)\Gamma(-\lambda+\mu+ika)}-
{\Gamma(1+2\lambda)\over \Gamma(\lambda+\mu+ika)\Gamma(1+\lambda-\mu+ika)}    
\over
{\Gamma(1+2\lambda)\over \Gamma(1+\lambda-\mu-ika)\Gamma(\lambda+\mu-ika)}-A{\Gamma(1-2\lambda)\over \Gamma(-\lambda+\mu-ika)\Gamma(1-\lambda-\mu-ika)}}~~,
\end{equation}
where $\Gamma(z)$ stands for the Gamma function with complex argument $z$, see e.g. \cite{[Ab65]}.

\begin{equation}
\label{A}
A=
\Big(\frac{b}{1+b}\Big)^{2\lambda}(1+b)^{-2ika}
\frac{_2F_1\big(\lambda+\mu+ika,1+\lambda-\mu+ika,1+2\lambda;\frac{b}{1+b}\big)}
{_2F_1\big(-\lambda+\mu-ika,1-\lambda-\mu-ika,1-2\lambda;\frac{b}{1+b}\big)}~,
\end{equation}
where $b=e^{-R/a}$ and
$_2F_1(a,b,c;z)$ stands for the Hyper-geometric Function  \cite{[Ab65]} with complex argument $z$.
The parameters have the following values:
\begin{equation}
\label{lam}
\lambda=ika\sqrt{1+\frac{V_1}{E}}~, \mu={1\over2}+{1\over2}\sqrt{1+4k^2a^2\frac{V_2}{E}}~, 
{\rm and}~E={\hbar^2 k^2\over 2M}~.
\end{equation}
The complex energy $E$ is calculated from the complex wave number $k$ by using the reduced mass $M$ of the target nucleus plus neutron system.
In order to find the poles of $S_0(k)$ in a domain of the complex $k$-plane we search for the zeros of $D(k)$, which is the denominator in Eq. (\ref{sform}):
\begin{equation}
\label{denom}
D(k)=
\frac{\Gamma(1+2\lambda)}{\Gamma(1+\lambda-\mu-ika)\Gamma(\lambda+\mu-ika)}-A\frac{\Gamma(1-2\lambda)}{\Gamma(-\lambda+\mu-ika)\Gamma(1-\lambda-\mu-ika)}~.
\end{equation}

The zero of $D(k)$ is found by the program BENCZE written  in Wolfram Mathematica.
Having the discrete set of zeros $k_m$ in the fourth $k$-quadrant
\begin{equation}
\label{zeros}
D(k_m)=0~,
\end{equation}
 we can order them as  the $k^R>0$ values increase.
The bound and the anti-bound poles along the imaginary $k$-axis are ordered differently, according to the number of the nodes $n$ in the radial wave function excluding the origin.

The  $a\to 0$ limit of the WS potential in Eq. (\ref{WS}) corresponds to the SQ potential. Poles  in the square well potential were studied extensively by Nussenzveig \cite{[Nu59]}. He showed, that for $l=0$ the pole trajectories converge to asymptotes with $k_m^R=\frac{m\pi}{R}$ as the depth $V_1\rightarrow 0$.
  The distance of the consecutive poles is determined basically by
the  radius of the square well $R$, where the wave function is reflected.
For large $m$ values the distance between two consecutive resonant poles:
$|k_{m+1}-k_m|$ approaches the value of $\frac{\pi}{R}$, so we can approximate 
the $m$ dependence of the pole position $k_m$ for large $m$ by
\begin{equation}
\label{rekn}
k^R_{m}= \frac{m\pi}{R} + O(1)~.
\end{equation}

In the book of R. Newton  \cite{[Ne82]} a similar expression is given for the real  parts of the starting values
of the $k$-trajectories.
The values there were given for so called strictly finite range (SFR) potentials vanishing  at and beyond $R_{max}$ \cite{[Da12]}. For a SQ potential the radius $R$ and the finite range $R_{max}$ is the same distance.  The WS or the GWS potentials  become zero only at infinite distance, therefore they are not  SFR type potentials. 
In Ref. \cite{[Sa14]} it was shown that for some types of the SFR potentials
the starting points of the pole trajectories (staring point of a trajectory is
a $k$ value belonging to very small potential strength $V_1$) can be described
by a relation:
\begin{equation}
\label{modkn}
|k_{m}|= \frac{m\pi}{R} + O(1)~.
\end{equation}
Although these findings was observed for very small $V_1$ values, we
speculate that Eq. (\ref{modkn}) might be valid approximately for realistic values of $V_1$, too.
 Therefore after calculating the complex $k_m$ eigenvalues with realistic values of $V_1$ we tried to fit the $|k_m|$ values  
by a first order polynomial: 
\begin{equation}
\label{line}
p(m)=a_0+a_1m~.
\end{equation}
The best fit first order polynomial minimizes the sum of the
squares of the differences:
\begin{equation}
\label{deltamod}
\Delta(a_0,a_1)=\sum_{m=m_s}^{m_u} [|k_{m}|-p(m)]^2 \to {\rm min}~.
\end{equation}
From the slope $a_1$ of the best fit polynomial
we can deduce a distance $\cal{R}$
based on the relation:
\begin{equation}
\label{range}
{\cal{R}}=\frac{\pi}{a_1}~. 
\end{equation} 
 Since the relation in Eq. (\ref{modkn}) expected to be valid for large $m$ values, we apply a lower cut $m_s$ for the $m$ values and check the validity of the linear behavior
as $m_s$ increases. The upper value of the index $m_u$ is fixed at a large value.

\section{Cut-off WS and GWS potentials}
By cutting off the WS potential or the GWS potential to zero at a finite $R_{max}$ distance 
we can convert them to SFR potential and  solve 
the radial equation by numerical integration.

At the direct numerical integration we proceed  from the origin $r=0$ step by step  to $R_{max}$ where  the nuclear potential becomes zero. At or beyond this distance, i.e. at $R_a\ge R_{max}$ we match the numerical solution to that of the free solution and calculate $S_0(k)$.
The cut-off  Woods-Saxon
potential (CWS) has the form:
\begin{equation}
\label{WSpot}
V^{CWS}(r)=V_1f^{CWS}(r)~,
\end{equation}
where the radial shape is
\begin{equation}
\label{vagottWS}
f^{CWS}(r)=-\left\{
\begin{array}{rl}
\frac{1}{1+e^{\frac{r-R}{a}}}
&\textrm{, if } r~<~R_{max}\\
0~~~~&\textrm{, if } r~\geq~ R_{max}~.
\end{array}
\right.
\end{equation}
In the cut-off GWS (CGWS) form  the surface potential term in Eq. (\ref{surfaceform})
is cut to zero at the same $R_{max}$ distance.

Although the method of the numerical calculation of the $k_m$ eigenvalues are given in several places, see e.g. Refs. \cite{[Ve82]},\cite{[Ix95]},
\cite{[Bar15]},
let us sketch briefly how the pole solutions of the radial equation are calculated numerically. 
We introduce  left and  right solutions of the radial equation in 
Eq. (\ref{radial}) and an intermediate distance $R_{id}$, which separates the left and
right regions.
A left solution satisfies the initial values: 
\begin{equation}
\label{leftzero}
u_{left}(0,k)=0,{\rm ~and} \quad\quad u_{left}^\prime(0,k)=1~,
\end{equation}
and it is defined in the $r\in [0,R_{id}]$ interval. We get it by integrating the  Eq. (\ref{radial}) numerically  from the origin until $R_{id}$, where we calculate the logarithmic
derivative of the left solution:
\begin{equation}
\label{leftlgder}
L_{left}(k,R_{id})=\frac{u_{left}^\prime(R_{id},k)}{u_{left}(R_{id},k)}~.
\end{equation}
The right solution satisfies outgoing boundary condition at the distance $R_a\ge R_{max}$, where the nuclear potential is zero, therefore the initial values for the right
solution are outgoing waves:
\begin{equation}
\label{righbc}
u_{right}(R_a,k)=e^{ikR_a},{\rm ~and} \quad\quad u_{right}^\prime(R_a,k)=ik e^{ikR_a}~.
\end{equation}
The right solution is defined in the $r\in [R_{id},R_a]$ interval. We integrate radial equation in Eq. (\ref{radial}) numerically in backward direction, starting from $R_a$ till $r=R_{id}$, where we calculate the logarithmic
derivative of the right solution:
\begin{equation}
\label{rightlgder}
L_{right}(k,R_{id})=\frac{u_{right}^\prime(R_{id},k)}{u_{right}(R_{id},k)}~.
\end{equation}
The  eigenvalue $k_m$  of the pole state belongs to the
zero of the difference of the left and right logarithmic derivatives:
\begin{equation}
\label{logder}
G(k_m,R_{id})=L_{left}(k_m,R_{id})-L_{right}(k_m,R_{id})=0~.
\end{equation}

The computer programs GAMOW \cite{[Ve82]}, and ANTI \cite{[Ix95]}  find the zeros of $G(k_m,R_{id})$, at certain $R_{id}$ matching distance $0<R_{id}<R_{max}\le R_{a}$.
The computer program GAMOW \cite{[Ve82]} uses Fox-Goodwin method with fix mesh size, and the program ANTI \cite{[Ix95]}
uses the more powerful Ixaru's method \cite{[Ix84]} for the numerical integration.
 For a broad resonance the proper choice of the $R_{id}$ matching distance is difficult. The zero  is searched by Newton iterations, and the iteration
process often converges poorly or fails. Therefore we developed a new method in which we compare the logarithmic derivatives not at a fixed $R_{id}$ distance but in a wide region in $r$. This method is built into the program JOZSO\footnote{The program name is chosen to honor   the late J\'ozsef Zim\'anyi to whom one of the authors (T. Vertse) is grateful for starting his carrier.} \cite{[No15]}.
  We calculate $G(k_m,r)$ in Eq. (\ref{logder})  at equidistant mesh-points with mesh size $h$
 at $r_j=j h$, $j\in [i_1,i_2]$.  The  mesh points are taken from a region where the nuclear potential falls most rapidly.
Then
we search for the absolute minimum of the function of two real variables 
$k^R$, and $k^I$:
 \begin{equation}
\label{minima1}
F(k^R,k^I)=\log [\sum_{j=i_1}^{i_2}  | G(k,r_j)|]~.
\end{equation}
 Absolute minima of the function ${F}(k^R,k^I)$ in Eq. (\ref{minima1})
should have a large negative value. The position of the absolute minimum is  the pole position of $S(k)$. The  minimum of the function is found by using the Powell's method in Ref. \cite{[numrec]}. The function $-F(k)$ shows peaks
at the minima of  $F(k)$.
  To find the minima
of the  function $F(k^R,k^I)$  first we explore the landscape of the
function $F(k)$  in a complex $k$ domain of our interest.
Then we search for the minima of the function ${F}(k^R,k^I)$ in Eq. (\ref{minima1}).

\section{Normalized resonant solution}

At the pole $k_m$ the left and the right solutions can be matched smoothly,
because their logarithmic derivatives are equal, or
we  take the left solution $u_{left}(r,k_m)$ in the interval $r\in [0,R_a]$ as a non-normalized solution of the radial equation in Eq. (\ref{radial}).
Sometimes we need the normalized solution, e.g. in Berggren-ensemble \cite{[Be68]}
in which
all pole solutions are normalized to unity.
The square of the norm is composed from the sum of the contributions of the internal and external regions:

\begin{equation}
\label{norm2}
N^2=N_{int}^2+N_{ext}^2~.
\end{equation}
The first one we calculate numerically, by quadrature
\begin{equation}
\label{normi}
N_{int}^2=\int_0^{R_a} u_{left}^2(r) dr~,
\end{equation}
while the second one is given analytically as in Ref. \cite{[Ve87]}.

\begin{equation}
\label{normext}
N_{ext}^2=-\frac{u_{left}(R_a,k_m)^2}{2ik_m}~.
\end{equation}
Then the normalized solution is simply:
\begin{equation}
\label{normsol}
u(r,k_m)=\frac{1}{N} u_{left}(r,k_m)~.
\end{equation}
Using the normalized solution we can  estimate
the energy shift $\Delta\epsilon_m$ of the pole energy $E$ of the CGWS potential
due to the change of the potential without cut. 
The tail of the resonant normalized solution beyond $R_a$ is an outgoing wave given as
\begin{equation}
\label{asol}
Ae^{ik_mr}=\frac{u(R_a,k_m)}{e^{ik_mR_a}}e^{ik_mr}~.
\end{equation}

We can try to correct the effect of the cut-off of the tail of the GWS potential on the energy of the resonance  using first order 
perturbation approach as:
\begin{equation}
\label{epert}
\Delta \epsilon_m=\int_{R_{max}}^\infty V^{GWS}(r) u^2(r,k_m) dr~,
\end{equation}
and compare it to the  energy difference
\begin{equation}
\label{ediff}
E({\rm BENCZE})-E({\rm JOZSO})~.
\end{equation}
If in the volume and in the surface terms of $V^{GWS}(r)$  in Eq. (\ref{general}) we approximate the fall of the  tails of the potentials  by $e^{\frac{R-r}{a}}$ in the integration region of Eq. (\ref{epert}), we can approximate the energy difference by the analytic expression
\begin{equation}
\label{apepert}
\Delta\epsilon_m=\frac{a(V_1+V_2)A^2}{(1-2ik_ma)}~e^{\frac{R-R_{max}}{a}+2ik_mR_{max}}~.
\end{equation}
A correction for the wavenumber shift $\Delta k_m$ can be calculated from the energy shift  $\Delta\epsilon_m$ by solving a second order algebraic equation

\begin{equation}
\label{kshift}
\Delta k_m=k_m+\sqrt{k_m^2+c_1 \Delta\epsilon_m}~,
\end{equation}
where $c_1$ denotes the factor between $k^2$ in fm$^{-2}$ and the energy $E$ in MeV ($k^2=c_1E$).
 The corrected resonance energy can be written as
\begin{equation}
\label{ecorr}
E^{corr}_m=E_m+\Delta\epsilon_m~,
\end{equation}
while the corrected wavenumber of the resonance has the form
\begin{equation}
\label{kcorr}
k^{corr}_m=k_m+\Delta k_m~.
\end{equation}
The accuracies of these corrections are checked by comparing the corrected  wave numbers to the values calculated by the program BENCZE, see Table \ref{sp56fe} of the next chapter.

If the perturbative correction  worked well for all poles for $l=0$ then we could use it later for $l>0$  when we 
are unable to handle the problem analytically.

\section{Numerical examples}

We applied our formalism for two systems in which neutrons are scattered on a heavy target nucleus and on a lighter nucleus. For the first one we choose the $^{208}$Pb+n system,
with potential parameters $V_1=44.4$ MeV, $V_2=0$, $r_0=1.27$ fm, $a=0.7$ fm.
The radius of the potential is $R=r_0~208^{1/3}\approx 7.52$ fm.

For the second example we considered a lighter system studied in Ref. \cite{[Ba15]}. In that work only bound states $^{56}$Fe+n system were calculated, here we extend the studies for  resonances.

\subsection{WS results for a heavy system}
\begin{figure}[h!]
\includegraphics[width=1.\columnwidth]{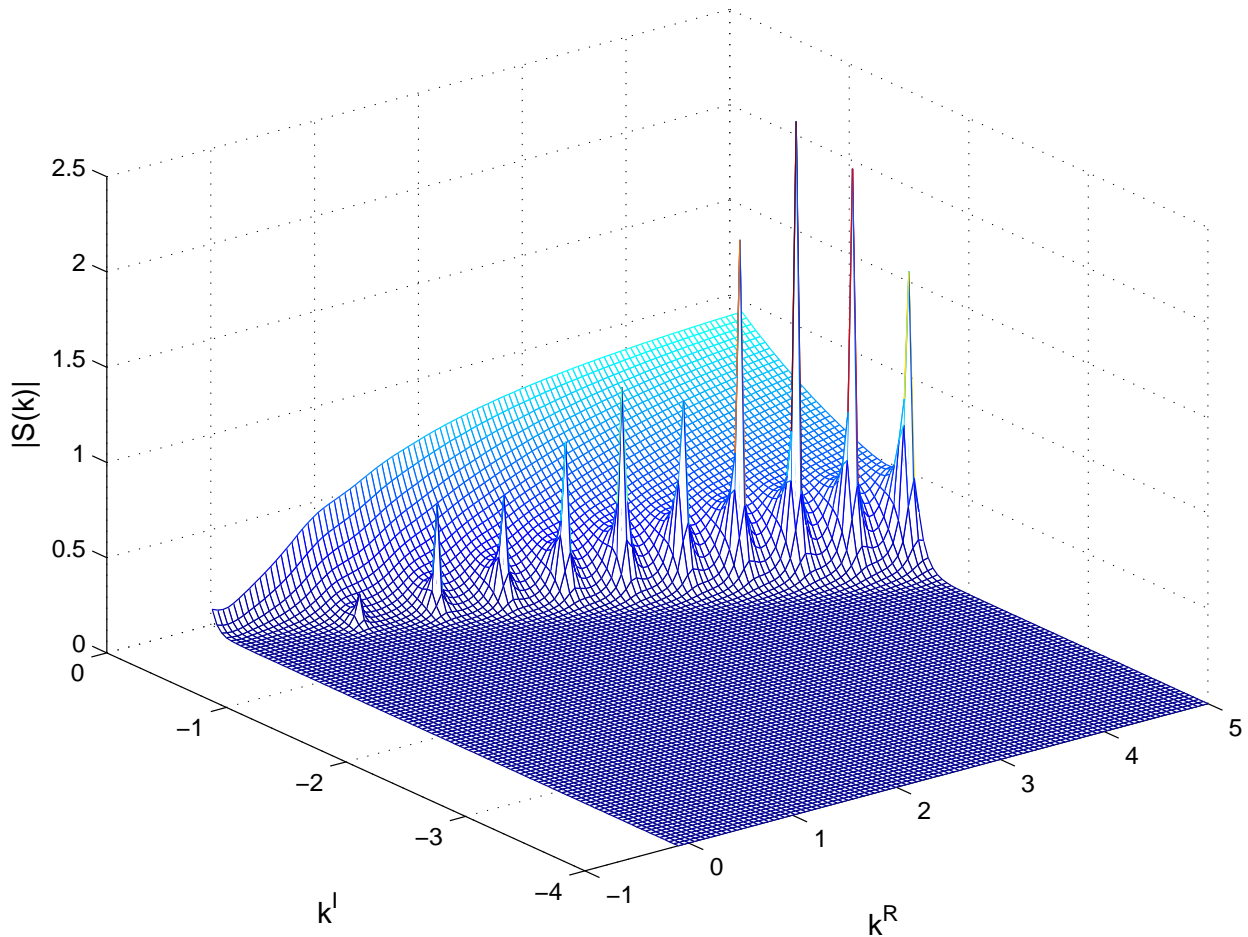}
\caption{ $|S(k)|$ on the domain
$k^R\in [-0.1,5]$ fm$^{-1}$ and $k^I\in [-4,-0.1]$ fm$^{-1}$ calculated for a WS
potential with depth parameter: $V_1=44.4$ MeV,  $V_2=0$ MeV, $a=0.7$ fm,
 $R=7.52$ fm }
\label{abssws}
\end{figure}

In Fig. \ref{abssws} we show the $|S_0(k)|$ on the domain
$k^R\in [-0.1,5]$ fm$^{-1}$ and $k^I\in [-4,-0.1]$ fm$^{-1}$ calculated for the WS
potential with the parameters listed above.
It can be seen that the poles form a mountain with peaks
being almost equidistant in $k^R$. If we assign a sequence number $m$ to each
peaks we can fit a first order polynomial in Eq. (\ref{line}) to either the
$k_m^R$ values as function of $m$, or to the $|k_m|$ values.
We observed that the fits to the
$k_m^R$ values and to the $|k_m|$ values produce very similar results.
Therefore in the remaining part of this paper we use only the fits to the $|k_m|$ values.
\begin{figure}[h!]
\includegraphics[width=1.\columnwidth]{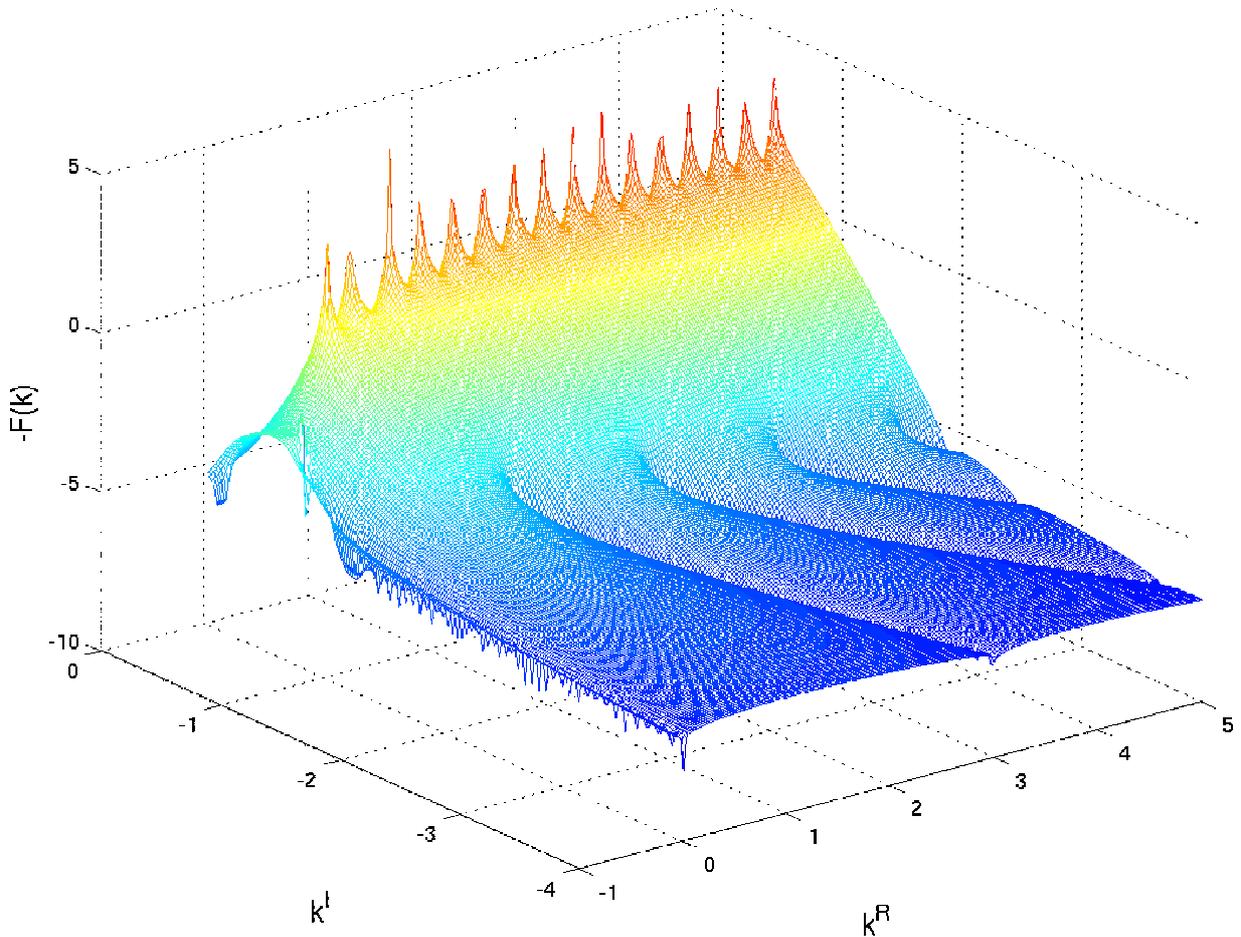}
\caption{ $-F(k)$ on the domain
$k^R\in [-0.1,5]$ fm$^{-1}$ and $k^I\in [-4,-0.1]$ fm$^{-1}$ calculated for a CWS
potential with depth parameter: $V_1=44.4$ MeV, $V_2=0$ MeV,  $a=0.7$ fm,
 $R=7.52$ fm, $R_{max}=12$ fm. }
\label{absscws}
\end{figure}

\begin{figure}[h!]
\includegraphics[width=1.\columnwidth]{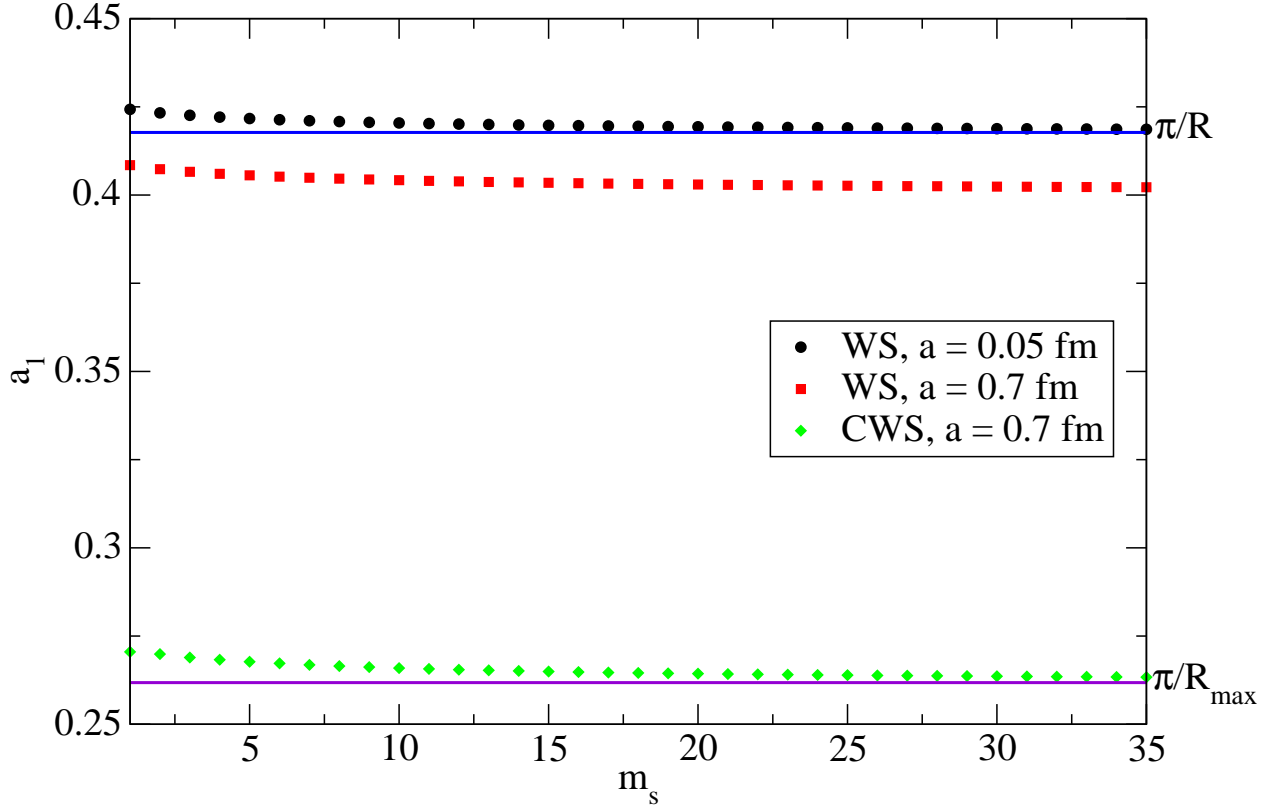}
\caption{Dependence of the slopes of the fitted lines on the lower cut value of the sequence number of the pole $m_s$ for the WS and CWS potentials, for $l=0$ and with
parameters: $V_1=44.4$ MeV,  $R=7.52$ fm
$a=0.7$ fm (red squares), $a=0.05$ fm (black circles). $R_{max}=12$ fm for CWS potential (green diamonds). The first order polynomials were fitted to the $|k_m|$ values. The horizontal lines correspond to the values $\pi/R$ (blue line) and $\pi/R_{max}$ (violet line). }
\label{varaslope}
\end{figure}

In Fig. \ref{varaslope} we show the $a$ dependence of the slope of the first order polynomial fitted to the  $|k_m|$ values calculated for the WS
potential with two different values of the diffuseness.
The very small, $a=0.05$ fm value simulates the SQ potential. The other case with $a=0.7$ fm is the normal diffuseness value for $^{208}$Pb.

The horizontal blue line corresponds to the value $\frac{\pi}{R}$, where $R$ is the common
radius of the square well and the WS potential. The distance of the asymptotes
of the poles in the SQ well is   $\frac{\pi}{R}$. 
As $m_s$ increases the $a_1$ slope values become soon independent on $m_s$ and their values are close to the $\frac{\pi}{R}$ value.
This simple property seems to be inherited from the SQ potential to the WS potential. The deviations from this simple rule increase  as the value of the diffuseness parameter
increases, but the deviations remain within $5\%$ of the $\frac{\pi}{R}$ value for $m_s>20$ even for $a=0.7$ fm (red squares). Therefore the estimated range in Eq. (\ref{range}) is close to the radius parameter of the WS potential.

\subsection{CWS results for a heavy system}
\label{heavy}
To compare to the cut-off potentials,
we can calculate the positions of the poles in a CWS potential with the same
parameters as the WS potential used, but with a cut-off radius $R_{max}=12$ fm.
The distributions of the complex poles can be visualized if we plot the landscape of the function $-F(k)$
defined in Ref. \cite{[Bar15]} on the same domain of the complex $k$-plane as we considered in Fig. \ref{abssws}. The results are displayed in Fig. \ref{absscws}.

The peaks of the  function $-F(k)$ are at the same $k$-values where the poles
of $S(k)$ are.
One can see in Fig. \ref{absscws} that the peaks form a single group of poles (mountain), but there are more peaks for CWS potential than in Fig. \ref{abssws} for the WS potential.

\begin{figure}[h!]
\includegraphics[width=1.\columnwidth]{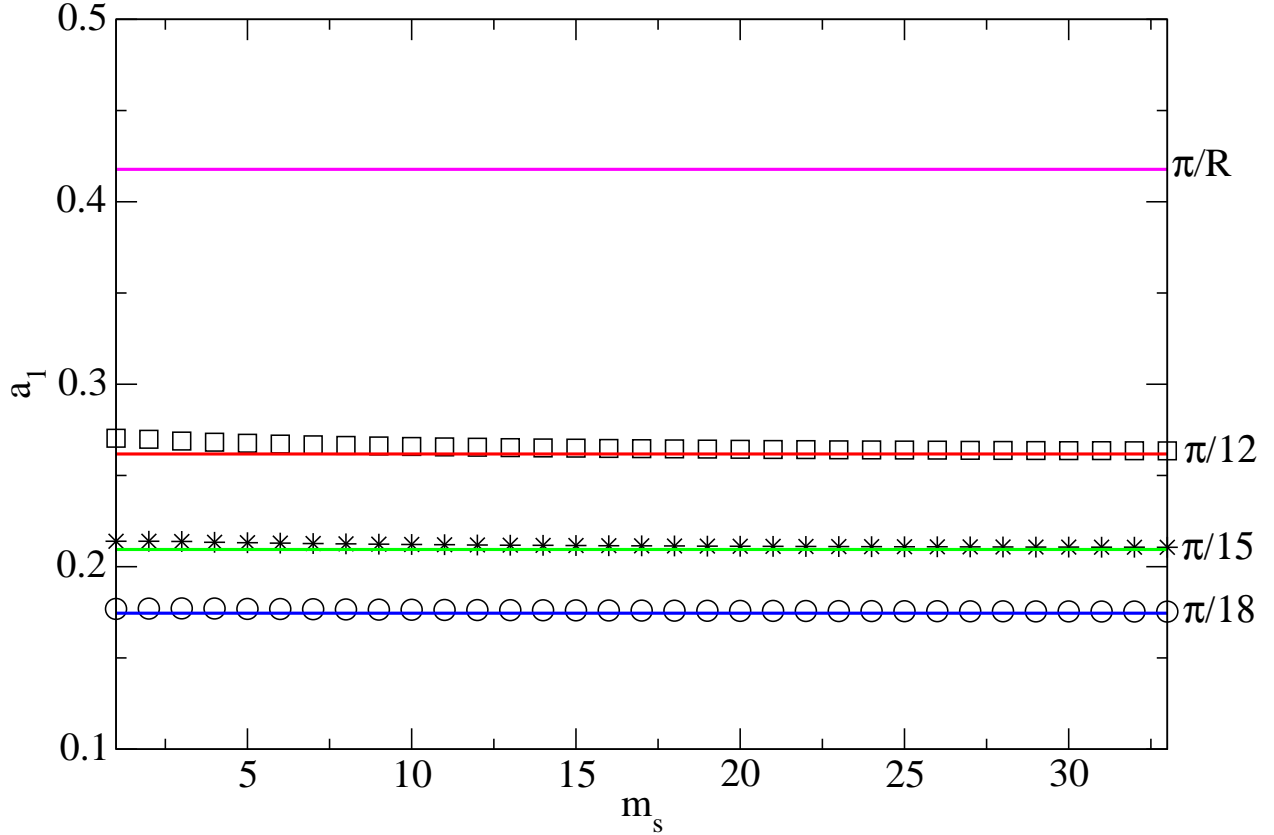}
\caption{Dependence of the slopes of the fitted lines on the lower cut value of the sequence number of the pole $m_s$ for the CWS potentials, for $l=0$ and with
parameters: $V_1=44.4$ MeV,  $R=7.52$ fm
$a=0.7$ fm  for different $R_{max}$ values. Squares for $R_{max}=12$, stars for
$R_{max}=15$, circles for $R_{max}=18$. The first order polynomials were fitted to the $|k_m|$ values. The horizontal lines corresponds to the values $\pi/R$ (magenta) and $\pi/R_{max}$, with $R_{max}=12$ (red line), $R_{max}=15$ (green line), $R_{max}=18$ (blue line). }
\label{varcutoffslope}
\end{figure}

It is interesting to study how the pole positions change as the diffuseness parameter $a$ of the CWS potential approaches zero to simulate a SQ potential.

As the value of the diffuseness is reduced  another mountain starts to develop at low
$k^R$ values. A similar phenomenon can be observed in the case of $^{56}$Fe in Fig. \ref{absscws2}.  The other mountain joins to the first mountain as $m$ increases (at higher energy). Reducing the diffuseness further the second mountain moves  away from the first one.

 It was observed in Ref. \cite{[Bar15]} that close lying resonances interact with each other, therefore 
we analyze the first mountain only when the other one is far enough not to interact with the resonances of the first mountain.

For a smoother CWS potential with  $a=0.7$ fm the reflection at $R$ is negligible and
the radial wave function is reflected only at $R_{max}$. Poles of the CWS in this case form a single mountain with slopes close to the value of $\pi/R_{max}$, see Fig. \ref{varaslope}. 
The $R_{max}$ dependence of the slope of the fitted first order polynomial are shown in Fig. \ref{varcutoffslope}. 
This dependence on the unphysical parameter $R_{max}$ is an  inconvenient feature of the CWS potential \cite{[ra11]}.

\subsection{CWS and CGWS potentials for a lighter system}

The lighter system is the $^{56}$Fe+n system studied by Bayrak and Aciksoz \cite{[Ba15]}. They used GWS and CGWS potentials for calculating bound state
energies in that system and found reasonable good agreements between the
bound state energies calculated by the analytical method and the numerical one
with cut-off potential. Now we extend the studies for resonant states and want to
investigate the effect of the cut-off on the positions of the resonant poles.
%
\begin{figure}[h!]
\includegraphics[width=1.\columnwidth]{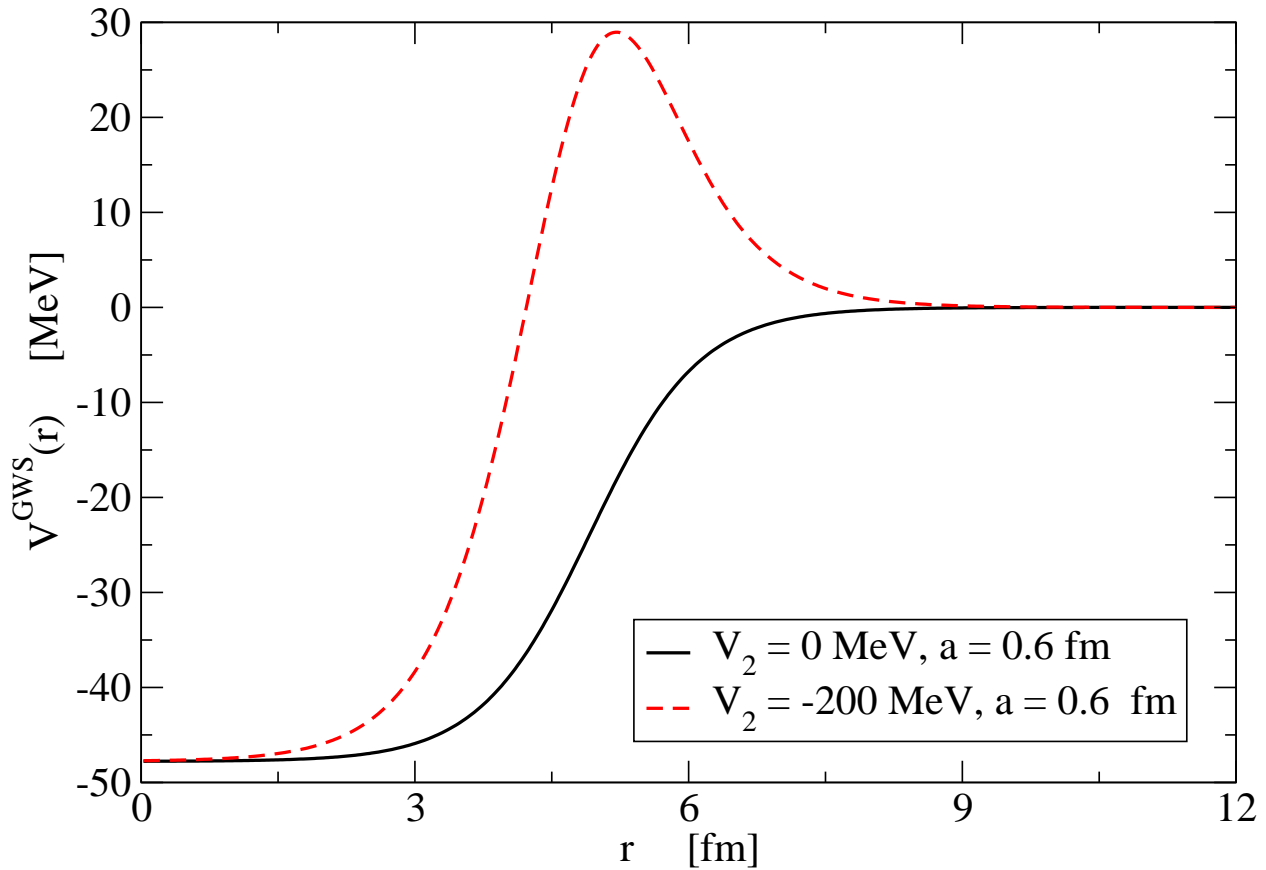}
\caption{ Radial shapes of the GWS potentials combined from
attractive volume and repulsive surface terms with parameters: $V_1=47.78$ MeV and $V_2=-200$ MeV,
$R=4.92$ fm, $a=0.6$ fm. }
\label{wsgws}
\end{figure}

In Fig. \ref{wsgws} one can see the radial shapes of 
the CGWS potential with strengths $V_1=47.78$ MeV and $V_2=-200$ MeV,
$R=4.92$ fm, $a=0.6$ fm. The parameters were taken from Bayrak \cite{[Ba15]}. Here the dashed red curve is the dot-dashed curve in Fig. 1 of that paper.
One can see in the figure that the  surface term with negative $V_2$ strength produces a barrier in
the GWS potential. The result of the barrier is the appearance of a few narrow resonances in that potential. If the barrier is high, then  there are more narrow
resonances. If we cut the tail of the GWS potential we introduce extra reflections
at the cut-off radius beside the ones at the nuclear radius $R$.

\begin{figure}[h!]
\includegraphics[width=1.\columnwidth]{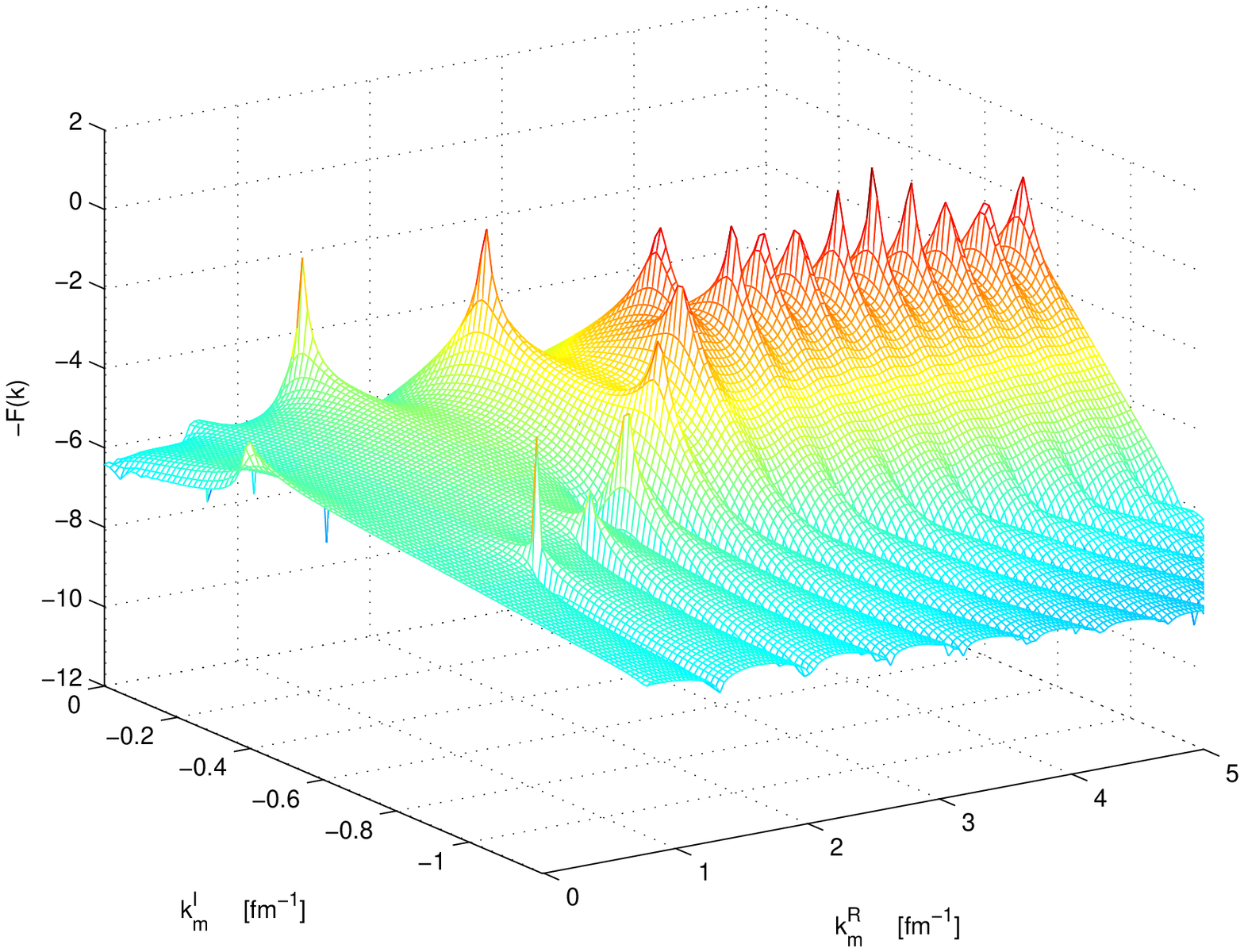}
\caption{ $-F(k)$ on the domain
$k^R\in [-0.1,5]$ fm$^{-1}$ and $k^I\in [-1.2,0.0]$ fm$^{-1}$ calculated for a CWS
potential with depth parameter: $V_1=47.78$ MeV, $V_2=-200$ MeV,  $a=0.6$ fm,
 $R=4.92$ fm, $R_{max}=12$ fm for $^{56}$Fe +n system.}
\label{absscws2}
\end{figure}

\begin{figure}[h!]
\includegraphics[width=1.\columnwidth]{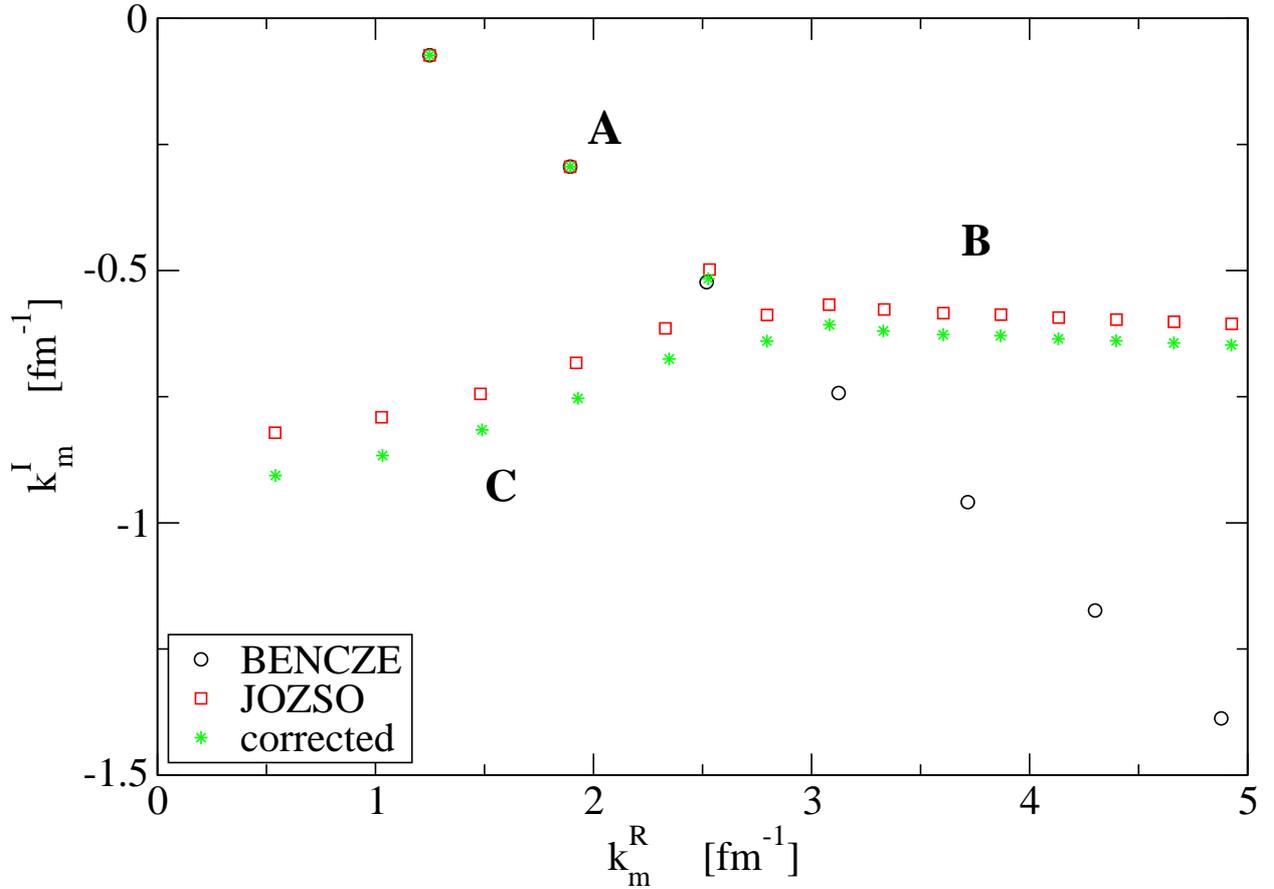}
\caption{ Positions of the poles in GWS and in CGWS potentials with
parameters $V_1=47.78$ MeV and $V_2=-200$ MeV,
$R=4.92$ fm, $a=0.6$ fm and $R_{max}=12$ fm without and with correction. }
\label{gwscgwsv2200}
\end{figure}

In Fig. \ref{absscws2} we plot the landscape of the function $-F(k)$ in the domain 
$k^R\in [-0.1,5]$ fm$^{-1}$, $k^I\in [-1.2,0]$ fm$^{-1}$ calculated by using CGWS potential. The positions of the poles for GWS and CGWS potentials are shown  in Fig. \ref{gwscgwsv2200}. Positions of the poles in GWS are denoted by black circles.

The poles of the GWS without cut  are
distributed quite regularly on the $k$-plane. If we fit the $m$-dependence of their $|k_m|$ values, the slope of the first order polynomial $a_1$ gives
a range value $\cal{R}$ being very close to $R=4.92$ fm. Therefore the reflection of the
wave function happens at the radius of the GWS potential.

 The poles for CGWS potential are denoted by red squares in Fig. \ref{gwscgwsv2200}. These poles were calculated by the program JOZSO. The results of the perturbation correction to the cut-off are denoted by green stars. Numerical values of the resonant pole positions are listed in Table \ref{sp56fe}.

\begin{table}[h!]
\begin{center}
\caption{Dependence of the complex $k_m$  wave numbers of the $l=0$ poles of the $S$-matrix  on the $V_2$ strength of the surface term  for $^{56}$Fe+$n$  in the GWS potential. Analytical results were calculated by the code BENCZE, the numerical results were calculated by the program {\rm JOZSO} \cite{[No15]}. The corrected $k_m^{corr}$ values were calculated using Eq. (\ref{kcorr}). The cut-off radius value was $R_{max}=12$ fm. The
strength of the volume term were kept fixed at $V_1=47.78$ MeV. $V_2$ is given in MeV, and all  wave numbers are in fm$^{-1}$ units.}
\label{sp56fe}
\begin{tabular}{ccccccc}
\hline\hline
$V_2$ & $k^R_m$(BENCZE) & $k^I_m$(BENCZE) & $k^R_m$(JOZSO) & $k^I_m$ (JOZSO) & ${\rm Re}(k_m^{corr})$ & ${\rm Im}(k_m^{corr})$\\
\hline\hline

$0$&$1.18982$&$-0.73207$&$1.15854$&$-0.61148$&$1.16310$&$-0.65090$\\
\hline
$-50$&$0.62716$&$-0.41298$&$0.62731$&$-0.41300$  &$0.62716 $&$ -0.41297$\\
$-50$&$1.56608$&$-0.62284$&$1.57202$&$-0.62082$&$1.56615$&$-0.62244$\\
$-50$&$2.26765$&$-0.83134$&$2.43836$&$-0.75586$&$2.42596$&$-0.78478$\\
\hline
$-100$&$0.94090$&$-0.21067$&$0.94088$&$-0.21071$&$ 0.94090$&$-0.21067$\\ 
$-100$&$1.70535$&$-0.47048$&$1.70194$& $-0.46605$& $ 1.70482$&$ -0.47034 $\\   
$-100$&$2.37343$&$-0.69445$&$2.40312$& $-0.63339$& $2.39838$&$-0.64374$\\   
\hline
$-200$&$1.24873$&$-0.07323$&$ 1.24873$& $-0.07323$&$1.24873$&$-0.07323$\\
$-200$&$1.89306$&$-0.29392$&$ 1.89311$& $-0.29429$&$ 1.89306$&$-0.29392 $\\
$-200$&$2.51849$&$-0.52297$&$ 2.53211$& $-0.49802$&$2.52623$&$-0.51614$\\

\hline\hline
\end{tabular}
\end{center}
\end{table}
The first line in Table \ref{sp56fe} shows one pole in WS and CWS potential
without barrier. There is no narrow resonance in this case. The resonance
in the CWS potential was selected as the closest pole to that of the WS potential shown in the first row. The correction of $k$ brings the pole in the CWS potential
somewhat closer to the one in the WS potential but the agreement is doubtful.

 The surface term with $V_2<0$ produces
a potential barrier.  As the height of the barrier increases, the differences between the wave numbers calculated by the program BENCZE and that of program
JOZSO, respectively become  much smaller than the difference for $V_2=0$ in the first row of the table.
The first and the second lines at each $V_2$ values show narrow resonances, where the CGWS results and their corrected values approach reasonably accurately
the GWS results in the second and third columns of the table.
The agreement for the third resonances is not so convincing therefore we
suspect that the given form of the first order perturbation correction in Eq. (\ref{epert}) does not work so well for
not narrow resonances. 
For the rest of the resonances not shown in the table, the differences of the
$k$ values in the potentials without cut and with cut-off are considerably larger.
The correction term  in Eq. (\ref{epert}) is clearly unable to correct the increasingly large differences originating from the cut-off the potential for broader resonances.

Three groups of poles can be observed in Fig. \ref{gwscgwsv2200}. The group A consists of the first three poles lying closest to 
the real axis. They  are the narrow resonances listed in the last three 
rows and fourth and fifth columns  of the Table \ref{sp56fe}. 
The positions of these narrow resonances approximate well the corresponding poles of the GWS potential calculated by the code Bencze. The results of the
correction improve the agreement even further. The general behaviour of these
group of resonances is similar to that of the resonances in GWS potential, namely that the distance of the resonances in group of A is determined by the 
radius $R$ of the CGWS potential. They are caused by the reflection of the radial wave function at the nuclear radius $R$. The A group of poles proceeds with another group (B) in which 
the distance of poles is much smaller and it is determined by the cut-off radius $R_{max}=12$  fm. The poles in group B 
are due to the reflection of the wave function at the cut-off radius. 
 The third group of poles 
(C) is composed of the five remaining broadest resonances in Fig. \ref{gwscgwsv2200}. 
We suspect that they are most probable due to the double reflections at $R$ and $R_{max}$. Their distance is determined approximately 
by the difference $R_{max}-R$, as one can see in Table \ref{range56fe}, where we varied the value of $R_{max}$. The largest difference is between $R_{max}-R$ and ${\cal R}$ is for $R_{max}=21$ fm. 
We suspect that it is due to the combined effect of the accumulation of numerical errors during the largest distance and the interaction 
with the closest poles in the other groups.

\begin{table}[h!]
\begin{center}
\caption{Comparism of the ranges calculated from the best fit first order polynomial in Eq. (\ref{range}) for the group of resonances B, A, C of the $^{56}$Fe+$n$ system  
in the CGWS potential with parameters: $V_1=47.78$ MeV, $V_2=-200$ MeV, $R=4.92$ fm, $a=0.6$ fm. Ranges ${\cal R}_A$, ${\cal R}_B$, ${\cal R}_C$ are 
the ranges corresponding to the groups A, B and C.}
\label{range56fe}
\begin{tabular}{cccccc}
\hline\hline
 $R_{max}$&${\cal R}_B~~~$&$R$&${\cal R}_A~~~$& $R_{max}-R$&${\cal R}_C$\\
\hline\hline
15   & 14.9999~~~~~&  4.92&4.8356~~~~~&  10.08 &    10.1911 \\
18   & 18.0360~~~~~&  4.92&4.7776~~~~~&  13.08 &    13.3406 \\
21   & 21.0829~~~~~&  4.92&4.7864~~~~~&  16.08 &    17.4930 \\
\hline\hline
\end{tabular}
\end{center}
\end{table}

The perturbation corrections in Eq. (\ref{kcorr}) are not large enough for the poles in group B and C to bring these poles to the vicinity of the poles calculated by the code BENCZE.  Therefore the correction works well only for the narrow resonances in group A.

The resonances in group A can be considered as physical resonances, since there
positions depend on the physical parameters of the GWS potential and they are
practically independent of the cut-off radius. The (in)dependence of the positions is shown in Table \ref{sprmax56fe}. The position of the $m=1$ resonance does not change when the value of the $R_{max}$ increased to $21$ fm
from $12$ fm. A similar increase of the $R_{max}$ value changed the positions
of $k_2$ and $k_3$ only in the last three decimal digits. So the dependence on the unphysical parameter can be neglected for this group of resonances.

\begin{table}[h!]
\begin{center}
\caption{Dependence of the complex $k_m$  wave numbers of the $l=0$ poles of the $S$-matrix  on the value of $R_{max}$
for the three narrow resonances of the $^{56}$Fe+$n$ system  in the CGWS potential with parameters: $V_1=47.78$ MeV, $V_2=-200$ MeV, $R=4.92$ fm, $a=0.6$ fm. }
\label{sprmax56fe}
\begin{tabular}{cccc}
\hline\hline
$m$& $R_{max}$& $k^R_m$& $k^I_m$\\
\hline\hline
$1$&$15.0$&$ 1.24873$& $-0.07323$\\
$1$&$18.0$&$ 1.24873$& $-0.07323$\\
$1$&$21.0$&$ 1.24873$& $-0.07323$\\
\hline
$2$&$15.0$&$ 1.89305$& $-0.29393$\\
$2$&$18.0$&$ 1.89306$& $-0.29392$\\
$2$&$21.0$&$ 1.89306$& $-0.29392$\\
\hline
$3$&$15.0$&$ 2.51152$& $-0.52311$\\
$3$&$18.0$&$ 2.51952$& $-0.52337$\\
$3$&$21.0$&$ 2.51840$& $-0.52283$\\
\hline\hline
\end{tabular}
\end{center}
\end{table}

 In Fig. \ref{gwscgwsv2200} the positions of the first three
resonances in group A are on the top of the resonances of the GWS potential. The
differences are small and can not be seen on the scale of the figure.
The corrected eigenvalues of these three resonances are on the top
of the resonances of the GWS potential, because the corrections are also
small for these narrow resonances.

For the resonances in group A the pole position is
practically independent of the cut-off radius
(Table \ref{sprmax56fe}), and the results of the programs
BENCZE and JOZSO are almost the same, as one can see   in Table \ref{sp56fe}. 
The number of the narrow resonances increases with the height of the barrier.

\begin{figure}[h!]
\includegraphics[width=1.\columnwidth]{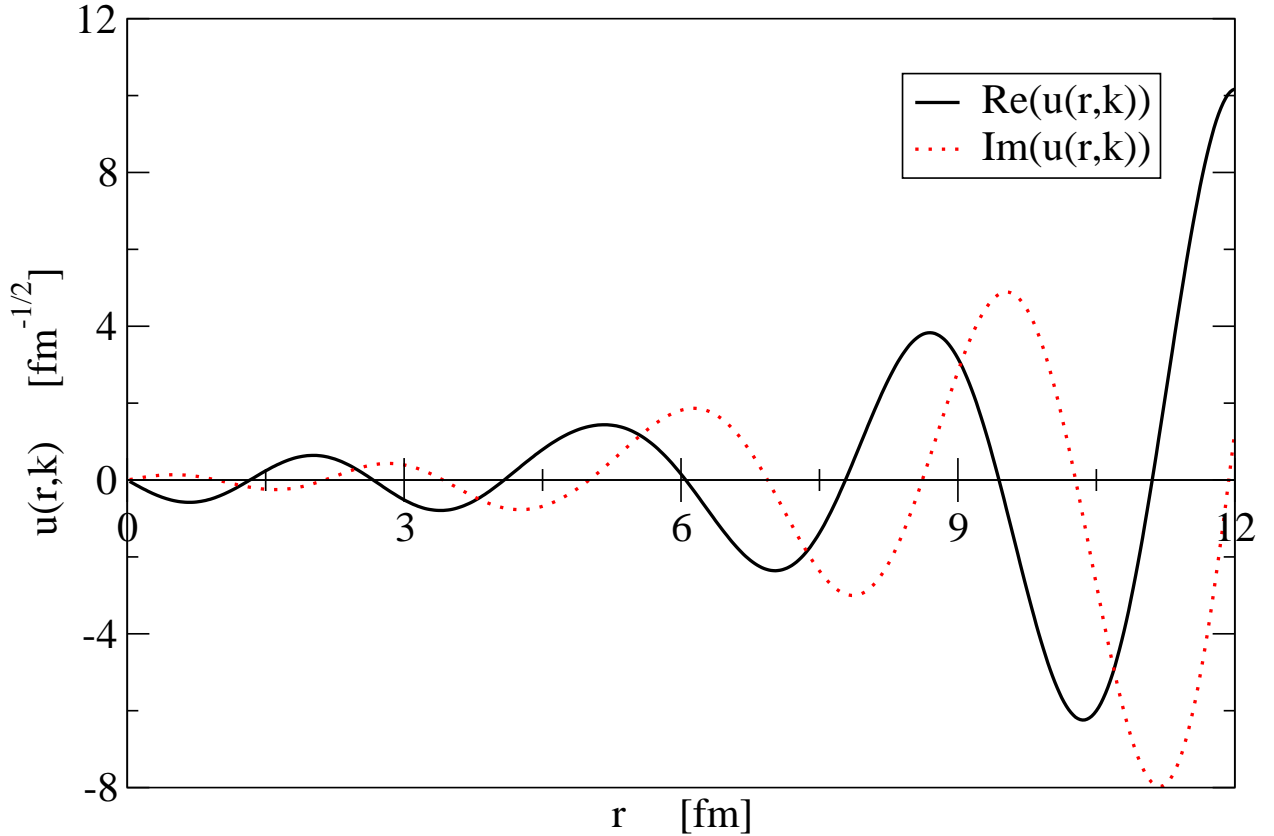}
\caption{ Radial shapes of the $m=2$ normalized resonant wave function with $k=(1.893,-0.294)$ fm$^{-1}$
in CGWS  potential with parameters: $V_1=47.78$ MeV and $V_2=-200$ MeV,
$R=4.92$ fm, $a=0.6$ fm, $R_{max}=12$ fm. }
\label{wfn3}
\end{figure}

Fig. \ref{wfn3} shows the normalized radial wave function for one of the narrow  resonances in CGWS potential with parameters: $V_1=47.78$ MeV and $V_2=-200$ MeV,
$R=4.92$ fm, $a=0.6$ fm and $R_{max}=12$ fm. It is the $m=2$ resonance in Table
 \ref{sprmax56fe} .
In the figure  the real and the imaginary parts of the
resonant wave function oscillate around the zero axis. In the external region
($r>R$) the magnitudes of the oscillations increase as $r$ increases. In the internal  region of the potential, the number of zeros of the
real part of the  wave function is $3$. If we make a similar plot for the $m=1$ resonant wave function   the number of zeros of the
real part of the  wave function is $2$. This is in agreement with the finding
in Ref. \cite{[Ba15]} in which the 0s and 1s states of the $^{56}$Fe are bound states. 

For narrow resonances the number of zeros of the real parts can be considered as the generalizations of the node number $n$
of the bound states. The imaginary part of the wave function also oscillates around the zero axis with exponentially growing amplitudes in the external region. Here the phase of the oscillations is
shifted wrt that of the real part. 
For this narrow resonance the magnitude of the internal oscillations is smaller
than that of the real part, and the crossing points of the zero axis are not far from
the crossing points of the real part. We observed that these features are general characteristics of the Gamow wave functions for narrow resonances.
 In certain respect the narrow resonances resemble to the bound states and they are sometimes
called as {\it quasi-stationary states}.
For the quasi-stationary states the corrections due to the cut-off the tail of the potential are small and their $k$ eigenvalues are not very sensitive to the
cut-off distance as we demonstrated it in Table \ref{sprmax56fe}.

 As the imaginary parts of $k$ become larger, the tail region moves inward, the imaginary part of the wave function competes with the real part. At the same time the value of the
cut-off distance becomes more important and the $k$ pole position of the CGWS potential
moves away from the $k$ of the GWS potential, as one can see in Fig. \ref{gwscgwsv2200}. The strong dependence on the cut-off distance was noticed
earlier by two of us in Ref. \cite{[Sa08]} for resonances with $l=5$.

\section{Summary}
We have investigated the effect of the cutting the tails of the WS and GWS potentials
for the resonant poles of the $S$-matrix by calculating the pole positions.
For zero angular momentum the radial Schroedinger equation is solved by using the analytical formula of Gy. Bencze \cite{[Be66]}. The positions of the poles
were calculated with high precision by the program BENCZE written in 
Wolfram Mathematica.
The cut-off versions of the same potentials, the CWS and CGWS potentials were
studied by solving the radial equation numerically by the FORTRAN program JOZSO \cite{[No15]}. Pole distributions were studied by exploring the landscape
of the modulus of the complex function $-F(k)$, which has maxima at the same
$k$ positions as the $S(k)$-matrix. The maxima of the function supplied us with
starting values for finding the accurate positions of the poles.
Presenting the landscapes of the $|S(k)|$ and the $|-F(k)|$ functions gives an
excellent tool to demonstrate the differences of the pole distributions of the
potentials without cut and with cut at finite distance.
With the absence of the surface term of the GWS potential we can study the normal WS and CWS potentials and the effect of the cut of their tails.

The pole structure of the WS and the CWS potential is basically different as it was pointed out by R. Newton \cite{[Ne82]}. The WS and the GWS potentials produce one
group of poles (mountain) similar to that of the square well potential with the same
nuclear radius. From the slopes of the first order polynomials fitted best to the moduli of the $k$ eigenvalues a range $\cal{R}$ can be deduced being very close to the nuclear radius parameter $R$. For these potentials the reflection
of the radial wave function takes place at $R$.

 The CWS and the CGWS potentials produce two or three groups of poles (mountains). They are more visible if in the CGWS potential a repulsive surface term is present and we have a potential barrier producing a few narrow resonances. 
The positions of the narrow resonances are similar to those of the GWS potential,
they form a first group of poles in which the distance of the poles are similar to that of the square well potential with radius $R$.

  The appearance of the second and sometimes the third group of poles  is apparently due to the cut of the CWS potential. The poles in the two different mountains are labelled
separately by indexing the order of their $k^R$ values. The moduli of the $|k_m|$ values in both mountains can be approximated by first order polynomials
as the function of the $m$ values. From the $a_1$ slopes of these polynomials one can
derive a sort of distance where reflection of the solution takes place. We assume that the appearance of the poles for broad resonances is due to reflections at these distances. Reflections take place when the derivative of the potential
has sudden change. In a square well this evidently happen at its radius $R$.
 For GWS potentials for small values of the diffuseness the derivatives are still large and the wave functions are reflected at the nuclear radius $R$.
 For a CGWS potential the potential has a jump at the cut-off radius $R_{max}$
 where the derivative does not exist. This causes a reflection of the wave function
 at this distance, and the distance of the poles in the first mountain of the
 diffuse well is influenced by this reflection. The $\cal{R}$ distances calculated from
 the slope of the best fit linear polynomials therefore are very close to the
 value of $R_{max}$.
 
 For small diffuseness and for not very large energy the radial wave function
 can be reflected at $R$ and at $R_{max}$ as well and oscillations between these
 two distances produces the third group of poles (mountain).
 
 The effect of cutting off the tail of the potential is too large for the second
 and the third group of poles and the first order perturbation correction is
 unable to recover the cut.
 
  The numerical results were performed for two different nuclear systems, the heavy target case for $^{208}$Pb+n system and for a lighter system for
  $^{56}$Fe+n system studied recently by Bayrak and Aciksoz \cite{[Ba15]}.
   We observed similar results for the
light and the heavy target systems, as far as the reflections of the wave function
are concerned and conclude that our results are typical
for the nuclear potentials studied.

\section*{Acknowledgement}

Authors are grateful to L. Gr. Ixaru and A. T. Kruppa for valuable discussions.
This work was  supported by the Hungarian Scientific Research -- OTKA Fund No. K112962.

\bibliographystyle{elsarticle-num}

\end{document}